\documentclass[reprint,amsmath,amssymb,aps,prd,showpacs,longbibliography,groupedaddres,
titlepage,nofootinbib]{revtex4-1}

\usepackage{graphicx}
\usepackage{dcolumn}
\usepackage{bm}
\usepackage{hyperref}
\usepackage{accents}
\usepackage{tensor}
\usepackage{comment}
\usepackage{lineno} 
\newcommand{\ovg}[1]{\stackrel{(\gamma)}{#1}}

\DeclareMathOperator{\sgn}{sgn}

\begin{document}

	\title{Revisiting the solution of the second-class constraints of the Holst action }

	\author{Merced Montesinos} 
	\email[]{merced@fis.cinvestav.mx}
	\affiliation{Departamento de F{\'{\i}}sica, Cinvestav, Avenida Instituto Polit{\'e}cnico Nacional 2508, San Pedro Zacatenco, 07360 Gustavo A. Madero, Ciudad de M{\'e}xico, Mexico}
	
	\author{Jorge Romero}
	\email[]{ljromero@fis.cinvestav.mx}
	\affiliation{Departamento de F{\'{\i}}sica, Cinvestav, Avenida Instituto Polit{\'e}cnico Nacional 2508, San Pedro Zacatenco, 07360 Gustavo A. Madero, Ciudad de M{\'e}xico, Mexico}
	
	\author{Mariano Celada}
	\email[]{mcelada@fis.cinvestav.mx}
	\affiliation{Departamento de F{\'{\i}}sica, Cinvestav, Avenida Instituto Polit{\'e}cnico Nacional 2508, San Pedro Zacatenco, 07360 Gustavo A. Madero, Ciudad de M{\'e}xico, Mexico}

	
	\date{\today}
	
	\begin{abstract}
	
		In this paper we revisit the nonmanifestly Lorentz-covariant canonical analysis of the Holst action with a cosmological constant. We take a viewpoint close to that of F. Cianfrani and G. Montani [\href{https://link.aps.org/doi/10.1103/PhysRevLett.102.091301}{Phys. Rev. Lett. {\bf 102}, 091301 (2009)}] and realize that the solution of the second-class constraints that the authors provide is incomplete, thus not accounting for the correct local dynamics of general relativity. We then mend their approach by adding the  missing degrees of freedom to the solution and give a complete description of the resulting theory, which preserves Lorentz invariance but turns out to be endowed with a noncanonical symplectic structure.~Later on and without resorting to any gauge condition, we perform a Darboux transformation to bring this theory into a canonical form. Finally, we show that in the time gauge both formulations, namely the noncanonical and the canonical ones, lead to the Ashtekar-Barbero variables.

	\end{abstract}
	
	\maketitle

	\section{Introduction}

		It is very well-known that the Lorentz-covariant canonical analysis of the Holst action for general relativity~\cite{Holst} features the presence of second-class constraints~\cite{Alexcqg1720,Barros} (this is true for the Palatini action as well~\cite{peld1994115}). Because the canonical quantization program requires us to get rid of them, there are in literature several ways of tackling the second-class constraints either by introducing the Dirac bracket~\cite{Alexcqg1720} or by solving them explicitly; the latter can be divided into into one approach without manifest Lorentz covariance~\cite{Barros} and those that manifestly preserve it~\cite{MontRomCel,*MontRomEscCel}.~Nevertheless, an important simplification occurs when adopting the so-called ``time gauge'', which reduces the local Lorentz symmetry to its subgroup $SO(3)$ and gives rise to the $\mathfrak{su}(2)$-valued Ashtekar-Barbero variables~\cite{Barbero,Holst}. Remarkably, these variables have established the foundations of what is known today as loop quantum gravity~\cite{rovelli2004quantum,AshLewcqg2115,thiemann2007modern,RovelliLR,*Rovellicag2815} and also carry information about the Immirzi parameter~\cite{Immirzicqg1410}, which drops out ``on shell'' in the Lagrangian theory but is present ``off shell'' in the gauge symmetry of the Holst action~\cite{RefGSGR} as well as in both the spectra of quantum observables~\cite{RoveSmolinNucP442,*AshtLewacqg14} and the black hole entropy~\cite{RovPRLt.77.3288,*AshBaezcqg105,*Meisscqg21,*Agulloetcprl100,*EngleNuiPRL105} derived within the loop approach.
		
		Despite the elegance and beauty entailed by the use of the Ashtekar-Barbero variables to describe the phase space of general relativity, it cannot be denied that this simplicity comes at the expense of sacrificing local Lorentz invariance through the use of the time gauge, and then one may wonder whether the dependence of the quantum theory on the Immirzi parameter could be a consequence of this gauge fixing and whether one can implement the loop approach without resorting to it. These questions are not new, and much of the work contained in~\cite{Alexcqg1720,AlexVassprd644,AlexLivprd674,Cianfraniprl,NouiSIGMA72011,Nouiprd84044002} undertakes these issues, although no definitive answer has been given yet. Our work, although utterly classical, also deals with the issue of Lorentz invariance in the Hamiltonian setting of general relativity and hence could be helpful for understanding the role of this symmetry in canonical quantum gravity.
		
		One of the first attempts at doing loop quantum gravity without imposing the time gauge is the work of Cianfrani and Montani~\cite{Cianfraniprl}, who provided a certain parametrization of the solution of the second-class constraints resulting in a noncanonical symplectic structure (for which they also asserted to have found canonical variables). Nevertheless, such a symplectic structure is incorrect because it involves 21 variables and thus does not give the right count of the local degrees of freedom (d.o.f.) of general relativity. What is wrong with their approach? It turns out that their solution of the second-class constraints is not the most general one as implied by these constraints. More precisely, the expression for the spatial part of the Lorentz connection provided by them involves only nine of the 12 free variables entailed by the homogeneous solution of the second-class constraints. Therefore, Cianfrani and Montani's solution is incomplete and thus not equivalent to the second-class constraints they began with. What is more, because of the missing free variables, the parametrization of the Lorentz connection given by Cianfrani and Montani cannot be related to the one given by Barros in Ref.~\cite{Barros} as they claim to do, for Barros actually provides a solution of the second-class constraints that ultimately leaves the two propagating d.o.f. of general relativity without performing any gauge fixing whatsoever. Hence, the procedure of Cianfrani and Montani is undoubtedly not right.
		
		In this work we mend the path followed by Ref.~\cite{Cianfraniprl}; we provide the most general solution of the second-class constraints in the same spirit of Ref.~\cite{Cianfraniprl} that correctly reproduces the local d.o.f. of general relativity, but that leads to a noncanonical symplectic structure. We then perform a Darboux transformation that clearly maps these variables to a canonical set similar to the one introduced in Ref.~\cite{Barros}; this is what we think Cianfrani and Montani should have done in order to establish an appropriate link between their work and the one by Barros. Finally, we show that, in the time gauge, the aforementioned Darboux map coincides with Barbero's canonical transformation up to a term proportional to the Gauss constraint, subsequently arriving at the Ashtekar-Barbero formulation, as expected.

	\section{Hamiltonian action with first- and second-class constraints}
		\label{HA}
		
		As usual in the canonical approach, we assume that the spacetime manifold $\mathcal{M}$ has topology $\mathcal{M}=\mathbb{R}\times\Sigma$, with $\Sigma$ being a spatial compact 3-manifold without a boundary. We use adapted coordinates $(t,x^a)$, where $t$ stands for the time direction and $\{x^a\}_{a=1}^3$ are coordinates on $\Sigma$, and represent time derivatives by a dot over the corresponding variable. Indices in capital latin letters ($I,\, J,\,\dots$) denote internal flat indices, which take values $\{0,i\}$, with  $i=1,\, 2,\, 3$; they are raised and lowered with the metric $\left(\eta_{IJ}\right) = \mbox{diag} \left(\sigma,1,1,1\right)$, for $\sigma=\pm 1$. We designate by $SO(\sigma)$ the internal group, where $SO(\sigma=-1)=SO(1,3)$ and $SO(\sigma=+1)=SO(4)$. Regardless of the nature of indices, we define the symmetrizer and the antisymmetrizer by $A_{(\alpha \beta)} := (A_{\alpha \beta} + A_{\beta \alpha})/2$ and  $A_{[\alpha \beta]} := (A_{\alpha \beta} - A_{\beta \alpha})/2$, respectively. For any quantity $U_{IJ}\ (= -U_{JI})$ taking values in the Lie algebra of $SO(\sigma)$, we define the endomorphism,
		\begin{equation}
			\ovg{U}_{IJ} := U_{IJ} + \dfrac{1}{2\gamma} \epsilon_{IJ}{}^{KL}U_{KL},
		\end{equation}
		where $\gamma\in\mathbb{R}$ ($\gamma\neq 0$ and $\gamma^2\neq\sigma$) is the Immirzi parameter and $\epsilon_{IJKL}$, with $\epsilon_{0123}=+1$, is totally antisymmetric (in three dimensions $\epsilon_{ijk}:= \epsilon_{0ijk}$).  When working with a tensor density we will indicate its positive (negative) weight with the number tildes above (under) it (for weights higher than +2 or lower than -1 the tildes will be omitted, but the weight will be specified elsewhere).
		
		In the Lorentz-covariant canonical analysis of the Holst action with a cosmological constant one arrives at the following~\cite{Barros} (see also Ref.~\cite{CelMont,*Romero2016} for the canonical analysis of its $BF$ counterpart):
		\begin{equation}
			\label{S}
			S=\int_\mathbb{R} dt \int_{\Sigma} d^{3}x\left( \ovg{\tilde{\Pi}} {}^{aIJ}\dot{\omega}_{aIJ}-\tilde{H}\right),
		\end{equation}
		where $(\omega_{aIJ}, \, \ovg{\tilde{\Pi}}{}^{aIJ} )$ [or equivalently $(\ovg{\omega}_{aIJ}, \, \tilde{\Pi}^{aIJ})$] are the canonical variables that parametrize the extended phase space at this level, and $\tilde{H}$ is the Hamiltonian density that turns out to be a linear combination of constraints,
		\begin{equation}
			\label{H}
			\tilde{H} =  \xi_{IJ}\tilde{\mathcal{G}}^{IJ} + N^{a}\tilde{\mathcal{V}}_{a} + \underaccent{\tilde}{N}\tilde{\tilde{\mathcal{H}}} + \underaccent{\tilde}{\varphi}{}_{ab}\tilde{\tilde{\Phi}}^{ab}+\psi_{ab}\Psi^{ab},
		\end{equation}
		with $\xi_{IJ}, \, \underaccent{\tilde}{N},\, N^{a}, \,  \underaccent{\tilde}{\varphi}_{ab}$ and $\psi_{ab}$ (of weight $-2$, since $\Psi^{ab}$ has weight $+3$) being Lagrange multipliers that enforce the constraints
		\begin{subequations}
			\label{const1}
			\begin{eqnarray}
				\label{Gauss}
				\tilde{\mathcal{G}}^{IJ} & := & D_{a} \stackrel{(\gamma)}{\tilde{\Pi}}\!\!{}^{aIJ} = \partial_{a} \ovg{\tilde{\Pi}}\hspace{-4pt}{}^{aIJ} + 2\omega_{a}{}^{[I|}{}_{K} \ovg{\tilde{\Pi}}\hspace{-4pt}{}^{aK|J]} \approx 0,\quad \\
				\label{Vector}
				\tilde{\mathcal{V}}_{a} & := & \frac{1}{2}\tilde{\Pi}^{bIJ}\stackrel{(\gamma)}{F}\!\!{}{_{baIJ}} \approx 0, \\
				\label{Scalar}
				\tilde{\tilde{\mathcal{H}}} & := & \frac{1}{2} \tilde{\Pi}^{aIK}\tilde{\Pi}^{b}{}_{K}{}^{J}\stackrel{(\gamma)}{F}\!\!{}{_{abIJ}} + \sigma\Lambda \tilde{\tilde{q}} \approx 0 ,\\
				\label{phi}
				\tilde{\tilde{\Phi}}^{ab} & := & -\sigma\epsilon_{IJKL}\tilde{\Pi}^{aIJ}\tilde{\Pi}^{bKL} \approx 0, \\
				\label{psi}
				\Psi^{ab} & := & \epsilon_{IJKL}\tilde{\Pi}^{(a|IM}\tilde{\Pi}^c{}_M{}^{J}D_c \tilde{\Pi}^{|b)KL}\approx 0,   	
			\end{eqnarray}
		\end{subequations}
 		where, $F_{abIJ} := 2 \left( \partial_{[a}\omega_{b]IJ} + \omega_{[a|IK} \omega_{|b]}{}^{K}{}_{J} \right)$ is the curvature of the spatial part of the $SO(\sigma)$ connection 1-form $\omega_{IJ}$, $\Lambda$ is the cosmological constant, and $\tilde{\tilde{q}}:=\det(q_{ab})>0$ is the determinant of the induced metric on $\Sigma$ (or spatial metric), whose inverse is determined by the relation $\tilde{\tilde{q}}q^{ab}= (\sigma/2) \tilde{\Pi}^{aIJ}\tilde{\Pi}^{b}{}_{IJ}$. Here, $\tilde{\mathcal{G}}^{IJ}, \, \tilde{\mathcal{V}}_{a},$ and $ \tilde{\tilde{\mathcal{H}}}$ (the Gauss, vector, and scalar constraints, respectively) are, in Dirac's terminology~\cite{Henneaux}, first class, and as such they are responsible for generating the gauge symmetries of general relativity [local $SO(\sigma)$ transformations and spacetime diffeomorphisms\footnote{The diffeomorphism constraint (which generates diffeomorphisms tangent to $\Sigma$) is given by the combination $\tilde{\mathcal{D}}_a:=\tilde{\mathcal{V}}_{a}+(1/2)\omega_{aIJ}\tilde{\mathcal{G}}^{IJ}$, which takes the form 
 			\begin{equation}
 				\tilde{\mathcal{D}}_a=\frac{1}{2}\left[\partial_b\left(\ovg{\tilde{\Pi}} {}^{bIJ}\omega_{aIJ}\right)-\ovg{\tilde{\Pi}} {}^{bIJ}\partial_a\omega_{bIJ}\right]. \label{DC}
 		\end{equation}} ]. On the other hand, $\tilde{\tilde{\Phi}}^{ab}$ and $\Psi^{ab}$ are second class, which will be dealt with in Sec.~\ref{Sol}.
		
		Before closing this section, it is worth convincing ourselves that the theory embodied in the constraints \eqref{Gauss}--\eqref{psi} propagates $(2 \times 18 - 2 \times 10 - 12)/2 = 2$ d.o.f., as it must be for general relativity.

	\section{Nonmanifestly Lorentz-covariant solution of the second-class constraints}
		\label{Sol}

		Since one of our goals is to compare our results with those of Cianfrani and Montani~\cite{Cianfraniprl}, we will solve the second-class constraints following an approach close to theirs, which, as shown in Sec.~\ref{CV}, allows us to make contact with a parametrization in terms of canonical variables similar to those given by Barros in Ref.~\cite{Barros}. To proceed, we first split the configuration and momentum variables into their ``electric'' and ``magnetic'' components, and pick the electric ones as part of the independent variables parametrizing the phase space once all the second-class constraints have been explicitly solved. The internal indices are in consequence raised and lowered with the three-dimensional Euclidean metric $\delta_{ij}$, which is the remnant of the original internal metric $\eta_{IJ}$. The solution of the constraint \eqref{phi} reads ({\it cf.}~Ref.~\cite{Barros})
		\begin{subequations}
			\label{solPhi}
			\begin{eqnarray}
				\label{solPhia}
				\tilde{\Pi}^{ai0} & =: & \tilde{E}^{ai}, \\
				\label{solPhib}
				\tilde{\Pi}^{aij} & = & 2 \chi^{[i|}\tilde{E}^{a|j]},
			\end{eqnarray}
		\end{subequations}
	    with $\chi^{i}$ being an arbitrary internal 3-vector. Using this parametrization, the spatial metric takes the form
		\begin{equation}
			q_{ab} = \epsilon|\tilde{\tilde{E}}| \Theta^{ij} \underaccent{\tilde}{E}_{ai}\underaccent{\tilde}{E}_{bj},\label{spamet}
		\end{equation}
		where we have assumed that $\tilde{\tilde{E}} := \det (\tilde{E}^{ai})$ is nonvanishing, $\underaccent{\tilde}{E}_{ai}$ denotes the inverse of $\tilde{E}^{ai}$, $\epsilon:=\sgn( 1+ \sigma \chi_i \chi^i)$, and 
		\begin{equation}
			\label{T}
			\Theta^{i}{}_{j} := \delta^{i}_{j}  + \sigma \chi^{i} \chi_{j}.
		\end{equation}
		Hence, the internal vector $\chi^{i}$ can be interpreted as an obstruction for $\underaccent{\tilde}{E}_{ai}$ to become an orthonormal (densitized) triad of the spatial metric. For the sake of future computations, it is also convenient to employ the quantity
		\begin{equation}
			\label{e}
			\eta^{i}{}_{j} := \left( 1+ \sigma \chi_{k} \chi^{k} \right) \delta^{i}_{j} - \sigma \chi^{i} \chi_{j},
		\end{equation}
		which is related to $\Theta_{ij}$ through the equality $\Theta_{ij}= (1+\sigma \chi_{k} \chi^{k})\left(\eta^{-1}\right)_{ij}$.	
		
		In order to solve the remaining second-class constraints~\eqref{psi}, we introduce
		the Levi-Civita connection $\Gamma$ compatible with $q_{ab}$ that satisfies $\nabla_a q_{bc}=0$ and $\Gamma^{a}{}_{bc} = \Gamma^{a}{}_{cb}$. We then define $\Gamma_a{}^i{}_j$ as the quantity fulfilling
		\begin{eqnarray}
			\label{CD}
			\partial_{a} \tilde{E}^{bi} + \Gamma^{b}{}_{ca}\tilde{E}^{ci} - \Gamma^{c}{}_{ca} \tilde{E}^{bi} + \Gamma_{a}{}^{i}{}_{j}\tilde{E}^{bj}  = 0.
		\end{eqnarray}
		Notice that this relation completely determines $\Gamma_a{}^i{}_j$ in terms of $(\tilde{E}^{ai},\chi_{i})$. $\Gamma_a{}^i{}_j$ is just a suitable object that allows us to write in a compact form the particular solution of the constraint~\eqref{psi} given in Eq.~\eqref{partsol} below. Although $\Gamma_a{}^i{}_j$ can be interpreted as the Levi-Civita connection with respect to the nonorthonormal frame defined by Eq.~\eqref{spamet}, this geometrical interpretation (and its consequences) will not be used here because we want to relate our results to those of Ref.~\cite{Barros} where such an interpretation is not put forward. Furthermore, observe that in general $\Gamma_{aij}$ is not antisymmetric in the internal indices, unless we impose the time gauge ($\chi^i=0$), in which case it becomes the Levi-Civita connection compatible with the orthonormal frame $|\tilde{\tilde{E}}|^{-1/2}\tilde{E}^{ai}$.
		
		To solve the constraint~\eqref{psi}, it is worth realizing that it defines a nonhomogeneous linear system of six equations for the 18 components of the connection $\omega_{aIJ}$. As a result, the solution for the connection must contain 12 arbitrary parameters coming from the homogeneous  solution of the system. Choosing nine of these parameters as the ``electric'' components $\omega_{a0i}$, the constraint~\eqref{psi} can be reinterpreted as a system of six equations for the nine ``magnetic'' components $\omega_{aij}$ whose solution takes the form
		\begin{eqnarray}
				\omega_{aij} = \Omega_{aij}+ 2\sigma \omega_{a0[i} \chi_{j]} - 2 \underaccent{\tilde}{E}_{a k} \Theta^{k}{}_{[i} \tilde{Y}_{j]},\label{solPsi}
		\end{eqnarray}
		where the quantities $\tilde{Y}^{i}$ account for the three remaining free parameters of the homogeneous solution and $\Omega_{aij}$ is a particular solution given by
		\begin{equation}
			\Omega_{aij}:= \Gamma_{a i k} \Theta^{k}{}_{j} - \sigma \left(\eta^{-1}\right)_{ij} \chi_{k}\partial_{a} \chi^{k}+ \sigma \chi_{j} \partial_{a}{\chi_{i}},\label{partsol}
		\end{equation}
		or, more explicitly,
		\begin{eqnarray}
			\Omega_{aij} = && \Theta_{[i|k}\tilde{E}^{b}{}_{|j]} \left( \partial_{b} \underaccent{\tilde}{E}_a{}^{k} -
			\partial_{a} \underaccent{\tilde}{E}_b{}^{k}\right)\notag\\
			&& -  \Theta_{[i|k}\tilde{E}^{b}{}_{|j]}\underaccent{\tilde}{E}_{a}{}^k
			\tilde{E}^{cl}\partial_{b}\underaccent{\tilde}{E}_{cl} +
			\Theta^{kl} \underaccent{\tilde}{E}_{ak} \tilde{E}^b{}_{[i|}\tilde{E}^c{}_{|j]} \partial_c \underaccent{\tilde}{E}_{bl} \notag \\
			& & - \underaccent{\tilde}{E}_a{}^k \tilde{E}^b{}_{[i|}\partial_b \Theta_{|j]k} - \sigma \chi_{[i|}\partial_a
			\chi_{|j]}.\label{solPsi1}
		\end{eqnarray}
		At this point we can compare with the solution of the constraint~\eqref{psi} provided by Cianfrani and Montani in Ref.~\cite{Cianfraniprl}: we immediately realize that their solution completely neglects the last term on the right-hand side of Eq.~\eqref{solPsi}, meaning that the parameters $\tilde{Y}^{i}$, which are part of the homogeneous solution of the constraint~\eqref{psi} as seen above, do not exist in the approach of Cianfrani and Montani. In consequence, their approach is not right simply because the phase-space parametrization employed by them after solving the second-class constraints is not the most general implied by~\eqref{psi}.
		
		With the solution of the second-class constraints successfully accomplished, we now have to express the action (and the constraints) in terms of the resulting phase-space variables. Substituting Eqs.~\eqref{solPhia}, \eqref{solPhib}, \eqref{solPsi} and \eqref{solPsi1} into the action \eqref{S}, after some algebra we get
		\begin{eqnarray}
			\label{S2}
			S & = & \int_\mathbb{R} dt \int_{\Sigma} d^{3}x \left( \mu_{ai} \dot{\tilde{E}}^{ai} +\tilde{\nu}_{i}\dot{\chi}^{i} + \tilde{\alpha}^{ai} \dot{\omega}_{a0i} +  \beta_{i} \dot{\tilde{Y}}^{i}  \right. \notag \\
				& &  \left.  - \tilde{H}' + \partial_{a}{\tilde{B}^{a}} \right), 
		\end{eqnarray}
 		with the functions $\mu_{ai}, \, \tilde{\nu}_{i}, \, \tilde{\alpha}^{ai}$, and $\beta_{i}$ being given by
		\begin{subequations}
			\label{mult}
	 		\begin{eqnarray}
	 			\mu_{ai} & := &  \eta_{ij} \partial_{a}\chi^{j} + \partial_{a}\chi_{i} + 2 \left( 1+\sigma \chi^{k}\chi_{k}\right) \underaccent{\tilde}{E}_{aj}\chi^{j} \tilde{Y}_{i} \notag \\
	 				& &    -  2\underaccent{\tilde}{E}_{aj}\Theta^{j}{}_{i} \chi^{k} \tilde{Y}_{k} - \tilde{E}^{bl}\chi_{l} \left[2\Theta^{k}{}_{i} \partial_{[a}\underaccent{\tilde}{E}_{b]k} \phantom{\chi^{|k)} }  \right.  \hspace{-1cm} \notag \\
		 			& & \left. - 2\Theta^{jk}\underaccent{\tilde}{E}_{aj}\tilde{E}^c{}_{i} \partial_{[b}\underaccent{\tilde}{E}_{c]k} +\Theta^{k}{}_{i} \underaccent{\tilde}{E}_{ak} \tilde{E}^{cm}\partial_{b} \underaccent{\tilde}{E}_{cm} \right.  \hspace{-1cm}  \notag \\
		 			& &  \left. - 2\sigma \underaccent{\tilde}{E}_{a}{}^{j} \chi_{(j|}\partial_{b}\chi_{|i)}  \phantom{\chi^{|k)} } \hspace{-0.6cm}\right] - \tilde{E}^{b}{}_{i} \chi_{k}\left[ 2 \Theta^{jk}  \partial_{[b}\underaccent{\tilde}{E}_{a]j} \phantom{\chi^{|k)} } \hspace{-1cm}\right. \notag \\
		 			& & \left.  - \Theta^{jk} \underaccent{\tilde}{E}_{aj} \tilde{E}^{cl}\partial_{b} \underaccent{\tilde}{E}_{cl}  +2 \sigma \underaccent{\tilde}{E}_{aj}\chi^{(j|}\partial_{b} \chi^{|k)} \right] \notag \\
		 			& & -\dfrac{2}{\gamma} \epsilon_{ijk} \underaccent{\tilde}{E}_{al} \Theta^{jl} \tilde{Y}^{k}, \label{mult1} \\
		 			\tilde{\nu}_{i} & := &  4 \sigma \tilde{E}^{a}{}_{i} \tilde{E}^{bj}\chi_{j}\chi^{k} \partial_{[a}\underaccent{\tilde}{E}_{b]k} - 2 \sigma \tilde{E}^{a}{}_{[i}\chi_{j]} \chi^{j} \tilde{E}^{bk}\partial_{a}\underaccent{\tilde}{E}_{bk}  \notag \\
		 			& &  + 4 \sigma \tilde{E}^{a}{}_{[i}\chi_{j]}\omega_{a0}{}^{j}  + 4\sigma \tilde{E}^{a}{}_{[i|}\chi^{j} \partial_{a} \chi_{|j]} + 4\sigma \chi^{j} \chi_{[i}\tilde{Y}_{j]}  \notag \\
		 			& &  - \dfrac{2\sigma}{\gamma}\epsilon_{ijk}\tilde{E}^{aj}\biggl[ \tilde{E}^{bk} \partial_{a}(\chi^l \underaccent{\tilde}{E}_{bl}) +\omega_{a0}{}^{k} \notag \\
		 			& &  - \frac{1}{2}\chi^{k}\tilde{E}^{bl}\partial_{a} \underaccent{\tilde}{E}_{bl} \biggr],\label{mult2}  \\
		 		\tilde{\alpha}^{ai} & := & - 2 \eta^{i}{}_{j} \tilde{E}^{aj}, \label{mult3}\\
		 		\beta_{i} & := & 4\chi_{i}.\label{mult4}
	 		\end{eqnarray}
 		\end{subequations} 	
 		Likewise, $\tilde{H}'$ stands for the Hamiltonian density formed by the linear combination of the first-class constraints $\tilde{\mathcal{G}}^{IJ}$, $\tilde{\mathcal{V}}_{a}$ and $\tilde{\tilde{\mathcal{H}}}$ (see below), whereas the boundary term $\tilde{B}^{a}$ is given by
 		\begin{eqnarray}
	 		\tilde{B}^{a} := - 2\dot{\tilde{E}}^{ai}\chi_{i} + \dfrac{1}{\gamma}\epsilon_{ijk}\left(  \Theta^{il}\underaccent{\tilde}{E}_{bl}\tilde{E}^{aj}\dot{\tilde{E}}^{bk} -\sigma \dot{\chi}^{i}\chi^{j} \tilde{E}^{ak} \right),\notag\\
	 		\label{bound}
 		\end{eqnarray}
 		which can be ignored in the present context since $\Sigma$ has no boundary, but that notwithstanding we have displayed here because it will be reabsorbed in the Darboux map given in Sec.~\ref{CV}.
 		
 		For the sake of simplicity, before introducing the explicit expressions of the constraints, let us define the variable (whose relation with $\Gamma_{aij}$ becomes clear once we adopt the time gauge),
 		\begin{equation}
 			\Upsilon_{ai} := \frac{1}{2} \epsilon_{ijk} \omega_{a}{}^{jk},
 		\end{equation}
 		together with the auxiliary matrices
 		\begin{eqnarray}
 		P_{ij} & := & \delta_{ij} + \frac{\sigma}{\gamma} \epsilon_{ijk}\chi^{k}, \\
 		Q_{ij} & := & \frac{1}{\gamma} \delta_{ij} + \epsilon_{ijk}\chi^{k}.
 		\end{eqnarray}
 		All of this allows us to write the (first-class) constraints as
 		\begin{subequations}
 			\label{const2}
	 		\begin{eqnarray}
	  			\tilde{\mathcal{G}}^{i}_{\mbox{{\tiny boost}}} & := & \tilde{\mathcal{G}}^{0i} = - \partial_{a}\left( P^{i}{}_{j} \tilde{E}^{aj} \right) - \Omega_{a}{}^{i}{}_{j}P^{j}{}_{l} \tilde{E}^{al} \notag \\
	  				& & + 2 \sigma \tilde{E}^{a[i} \omega_{a0}{}^{j]}\chi_{j} + \frac{\sigma}{\gamma}\epsilon^{ijk} \omega_{a0j} \tilde{E}^{a}{}_{k}  \notag \\
	  				& & + \frac{1}{\gamma} \epsilon_{jkl} \omega_{a0}{}^{j}\tilde{E}^{ak}\chi^{l}\chi^{i} - \left( \eta^{i}{}_{j} + P^{i}{}_{j}\right)\tilde{Y}^{j},\label{Gboost}\\
	  			\tilde{\mathcal{G}}^{i}_{\mbox{{\tiny rot}}} & := & \frac{1}{2} \epsilon^{ijk} \tilde{\mathcal{G}}_{jk}  =  -\partial_{a}\left( Q^{i}{}_{j} \tilde{E}^{aj} \right) - \Omega_{a}{}^{i}{}_{j}Q^{j}{}_{l} \tilde{E}^{al} \notag \\
		  			& & + 2 \frac{\sigma}{\gamma} \tilde{E}^{a[i} \omega_{a0}{}^{j]}\chi_{j} + \epsilon^{ijk}  \omega_{a0j}\tilde{E}^{a}{}_{k} \notag \\
		  			& & + \sigma \epsilon_{jkl} \omega_{a0}{}^{j}\tilde{E}^{ak}\chi^{l}\chi^{i} - \left( \dfrac{1}{\gamma}\eta^{i}{}_{j} + Q^{i}{}_{j}\right)\tilde{Y}^{j},\label{Grot}\\
	 			\tilde{\mathcal{V}}_{a} & = &  - \omega_{a0i} \partial_{b}\left( P^{i}{}_{j} \tilde{E}^{bj} \right) - \Upsilon_{ai} \partial_{b}\left( Q^{i}{}_{j} \tilde{E}^{bj}\right) \notag \\
	 				& & +2 P^{i}{}_{j} \tilde{E}^{bj}\partial_{[a} \omega_{b]0i} + 2 Q^{i}{}_{j}\tilde{E}^{bj} \partial_{[a}\Upsilon_{b]i} \notag \\
	 				& & - \omega_{a0i}\tilde{\mathcal{G}}^{i}_{\mbox{{\tiny boost}}} - \Upsilon_{ai}\tilde{\mathcal{G}}^{i}_{\mbox{{\tiny rot}}},  \label{vect}\\
 				\tilde{\tilde{\mathcal{H}}} & = &  -\tilde{E}^{ai} \chi_{i}\tilde{\mathcal{V}}_{a}  
 				-\sigma  \left(1+\sigma \chi_{n}\chi^{n}\right) \epsilon_{ijk} \tilde{E}^{ai}
 				\tilde{E}^{bj} \notag \\
 				& & \times \left[  \dfrac{\sigma}{\gamma} \partial_{a}\omega_{b0}{}^{k} +
 				\partial_{a}\Upsilon_{b}{}^{k} - \dfrac{1}{2}\epsilon^{klm} \left(
 				2\dfrac{\sigma}{\gamma} \omega_{a0l}\Upsilon_{bm}  \right. \right. \notag \\
 				& & \left. \left. \phantom{\dfrac{1}{2}} \hspace{-0.5cm} + \sigma
 				\omega_{a0l}\omega_{b0m} +  \Upsilon_{al}\Upsilon_{bm}\right) \right] \notag \\
 				& &  +\sigma\Lambda\left|1+\sigma \chi_{i}\chi^{i}\right| |\tilde{\tilde{E}}|,\label{scal}
	 		\end{eqnarray}
 		\end{subequations}
	 	where we have split the Gauss constraint into its ``boost'' and ``rotational'' parts. \\
 		
 		From expressions \eqref{mult1}--\eqref{mult4}, we observe that $\mu_{ai}, \, \tilde{\nu}_{i}, \, \tilde{\alpha}^{ai}$, and $\beta_{i}$ are functions solely of the 24 phase-space variables ($\tilde{E}^{ai}, \, \chi_{i}, \, \omega_{a0i}, \,\tilde{Y}^{i}$), therefore giving rise to the symplectic potential. In particular, neither of the pairs $(\omega_{a0i},\tilde{E}^{ai})$ and $(\chi_{i},\tilde{Y}^{i})$ are canonical, which may complicate the construction of the quantum theory emanating from this phase-space parametrization, but at least classically these 24 variables subject to the ten first-class constraints \eqref{Gboost}--\eqref{scal} give an alternative and faithful description of the two propagating d.o.f. of general relativity. On the other hand, in the approach of Ref.~\cite{Cianfraniprl} the phase space is described just by 21 variables (an odd number), which translates in an incorrect number of local d.o.f. for the underlying theory. This inconsistency results from Cianfrani and Montani's solution of the system of equations defined by the constraint~\eqref{psi}, which actually corresponds to a particular solution of it and not to its most general solution; thus, not accounting for the homogeneous solution constitutes an incomplete parametrization of the phase space. What is more, even if we restrict our attention to the case when $\chi^i$ is time independent, as Cianfrani and Montani do in the first part of their analysis, it does not mean that the variables $\tilde{Y}^{i}$ do not exist anymore, but that they take some determinate values (nonvanishing in general) according to the compatibility of this assumption with the Gauss constraints \eqref{Gboost}--\eqref{Grot}, which in principle allows us to express these variables in terms of the remaining variables (fixing $\chi^i$ can also be interpreted as a gauge condition). Therefore, to assure that the resulting theory can also be cast as an $SU(2)$ gauge theory, the terms proportional to (fixed) $\tilde{Y}^{i}$ must also be carefully handled all along, something completely overlooked in Ref.~\cite{Cianfraniprl}.

	\section{Darboux map}
		\label{CV}
		
		As seen in the previous section, the phase space is parametrized by 24 noncanonical coordinates. Nevertheless, we can in principle construct, by means of a Darboux transformation, a set of canonical variables to account for the same kinematic d.o.f. In this section we explicitly exhibit this map, which in turn establishes a bridge between the results of Sec.~\ref{Sol} and those of Ref.~\cite{Barros}.\footnote{The variables defined in this paper are related with those introduced by Barros through the simultaneous changes $\tilde{E}^{ai}\rightarrow -\tilde{E}^{ai}$, $A_{ai} \rightarrow - \gamma A_{ai}$, and $\zeta_{i} \rightarrow \gamma \zeta_i$. This discrepancy with Barros allows us to make contact with the Ashtekar-Barbero formalism within usual conventions (see for instance Ref.~\cite{thiemann2007modern}).} We are able to accomplish this by keeping $\tilde{E}^{ai}$ and $\chi^{i}$ unchanged, while making the following definitions:
		\begin{subequations}
			\begin{eqnarray}
			A_{ai} & := & -\gamma \ovg{\omega}_{a0i} - \gamma \ovg{\omega}_{aij}\chi^{j}, \label{transf1}\\
			\tilde{\zeta}_{i} & := & -\gamma \ovg{\omega}_{aij}\tilde{E}^{aj}, \label{transf2}
			\end{eqnarray}
		\end{subequations}
		where the expression for $\omega_{aij}$ in terms of the noncanonical phase space variables is given in Eq.~\eqref{solPsi}. With these definitions we will replace both $\omega_{a0i}$ and $\tilde{Y}^{i}$ with $A_{ai}$ and $\tilde{\zeta}_{i}$. To express the action~\eqref{S2} in terms of the new variables, we first invert the relations \eqref{transf1}--\eqref{transf2}, giving
		\begin{subequations}
			\label{transfinv}
			\begin{eqnarray}
			\label{transfw}
			\omega_{a0i} &=& \left( \eta^{-1} \right)_{i}{}^{j} \bigg\lbrace-\frac{1}{\gamma} A_{aj}  -\frac{1}{2}\epsilon^{klm}Q_{kj}\Omega_{alm} \nonumber \\ 
			& & -\frac{\gamma^2}{2\left(\gamma^2\!-\sigma\!\right)} M_{jkl}\underaccent{\tilde}{E}_a{}^k \Theta^{l}{}_{m}\!\bigg[ \frac{1}{\gamma} \tilde{\zeta}^{m} - \tilde{E}^{b}{}_{n} \bigg( \frac{1}{\gamma}S^{mnp}A_{bp}  \notag \\
			& &  - T^{mnpq}\Omega_{bpq} \bigg) \bigg] \bigg\rbrace, \\
 			\tilde{Y}^{i}\! &=& \! -\frac{\gamma^2}{2\left(\gamma^2\!-\sigma\!\right)}\Theta^{i}{}_{j}\!\bigg[  \tilde{E}^{a}{}_{k} \bigg(T^{jklm}\Omega_{alm} - \frac{1}{\gamma}S^{jkl}A_{al} \bigg)\notag \\
 			& & +\frac{1}{\gamma} \tilde{\zeta}^{j}   \bigg], \nonumber \\ \label{transfY}
			\end{eqnarray}
		\end{subequations}
		where we have defined the following internal quantities:
		\begin{subequations}
			\label{Mdef}
			\begin{eqnarray}
			M_{ijk} & := &  \delta_{i j} \chi_{k} - \eta_{i k} \chi_{j} + \dfrac{1}{\gamma}\epsilon_{ijk} - \frac{\sigma}{\gamma} \epsilon_{i k l}\chi^{l}\chi_{j}, \\
			S_{ijk} & := &  \sigma\epsilon_{i j l} Q^{l m}  \left( \eta^{-1}\right)_{m k}, \\
			\label{Tdef}
			T^{ij}{}_{kl} & := & \delta^{i}_{[k} \delta^{j}_{l]}  - \frac{\sigma}{2}  \epsilon^{i j m} \epsilon_{k l q} \left(\eta^{-1}\right)^{n p} Q_{m n} Q^{q}{}_{p}.
			\end{eqnarray}
		\end{subequations}
		
		Substituting \eqref{transfw}--\eqref{Tdef} into \eqref{S2} is a lengthy but straightforward computation that results in
		\begin{equation}
		\label{S3}
		S = \int_\mathbb{R} dt \int_{\Sigma} d^{3}x \left( \frac{2}{\gamma} \dot{A}_{ai} \tilde{E}^{ai} + \frac{2}{\gamma}\dot{\chi}_{i}\tilde{\zeta}^{i} - \tilde{H}' \right),
		\end{equation}
		which tells us that the pairs $(A_{ai},\, \tilde{E}^{ai})$ and $(\chi_{i},\,\tilde{\zeta}^{i})$ are canonical, and that in consequence they obey the fundamental Poisson brackets $\{ A_{ai}(x),\, \tilde{E}^{bj}(y)\}=(\gamma /2)\delta^{b}_{a}\delta^{j}_{i}\delta^{3}(x-y)$ and $\{\chi_{i}(x),\,\tilde{\zeta}^{j}(y)\}=(\gamma /2) \delta^{j}_{i}\delta^{3}(x-y)$. In addition, the constraints \eqref{Gboost}--\eqref{scal} now read
		\begin{subequations}
			\label{const3}
			\begin{eqnarray}
				\tilde{\mathcal{G}}^{i}_{\mbox{{\tiny boost}}} & = & -\partial_{a}\left( P^{i}{}_{j} \tilde{E}^{aj} \right) + \dfrac{2\sigma}{\gamma} A_{aj} \tilde{E}^{a[j}\chi^{i]}+\dfrac{\sigma}{\gamma}\tilde{\zeta}_{j}\chi^{j}\chi^{i} \notag \\
					& & +\frac{1}{\gamma}\tilde{\zeta}^{i}, \label{boostCV}\\
				\tilde{\mathcal{G}}^{i}_{\mbox{{\tiny rot}}} & = & - \partial_{a}\left( Q^{i}{}_{j} \tilde{E}^{aj} \right) - \dfrac{1}{\gamma} \epsilon^{i}{}_{jk}\left(A_{a}{}^{j}\tilde{E}^{ak} -\tilde{\zeta}^{j}\chi^{k}\right),\label{gaussCV} \\
				\tilde{\mathcal{V}}_{a} & = & \dfrac{2}{\gamma}\tilde{E}^{bi}\partial_{[b}A_{a]i}- \dfrac{1}{\gamma}\tilde{\zeta}_{i}\partial_{a} \chi^{i} + \frac{\gamma^{2}}{\gamma^{2}-\sigma} \left[ \phantom{\frac{1}{2}} \hspace{-0.3cm} \left( Q^{i}{}_{j}\tilde{\mathcal{G}}^{j}_{\mbox{{\scriptsize boost}}}  \right. \right. \notag \\
					& &  \left.- P^{i}{}_{j}\tilde{\mathcal{G}}^{j}_{\mbox{{\scriptsize rot}}}\right)J_{a i} -  \dfrac{2\sigma}{\gamma^{2}} \tilde{E}^{b[i}\chi^{j]}A_{ai}A_{bj} \notag \\
					& &   + \dfrac{1}{\gamma^{2}}A_{ai}\left( \tilde{\zeta}^{i} +\sigma \tilde{\zeta}_{j}\chi^{j}\chi^{i}\right) - \frac{\sigma}{\gamma^{3}}\epsilon_{ijk}\left( \tilde{E}^{bi}A_{b}{}^{j} \phantom{\dfrac{1}{2}} \right.\notag \\
					& & \left. \left. \phantom{\frac{1}{2}} \hspace{-0.3cm} +\tilde{\zeta}^{i}\chi^{j} \right) A_{a}{}^{k} \right], \label{vecCV}\\ 
				\tilde{\tilde{\mathcal{H}}} & = &   -\tilde{E}^{ai}\chi_{i}\tilde{\mathcal{V}}_{a} 
				-\sigma \left(1+\sigma \chi_{p}\chi^{p}\right)
				\epsilon_{ijk}\tilde{E}^{ai}\tilde{E}^{bj} \left\lbrace \partial_{a}J_{b}{}^{k}
				\phantom{\dfrac{1}{2}} \right. \notag \\
				& & - \frac{\sigma \gamma^{2}}{2(\gamma^{2}- \sigma)} \left[ \epsilon^{klm} \left( \dfrac{1}{\gamma^{2}}
				A_{al}A_{bm} + \sigma J_{al}J_{bm}\phantom{\dfrac{1}{2}} \right.\right. \notag \\
				& & \left.  + \frac{2}{\gamma^{2}} A_{al}J_{bm} \right)  - \dfrac{2}{\gamma} \left( A_{al} + J_{al} \right) J_{b}{}^{k} \chi^{l} \notag \\ 
				& &  \left. \left. + \dfrac{2}{\gamma}A_{al} J_{b}{}^{l}\chi^{k}  +
				\epsilon^{lmn}J_{al}J_{bm}\chi_{n}\chi^{k} \phantom{\dfrac{1}{2}}
				\hspace{-0.25cm}\right] \right\rbrace \notag \\
				& & + \sigma\Lambda\left|1+\sigma \chi_{i}\chi^{i}\right| |\tilde{\tilde{E}}| .\label{scal1}
			\end{eqnarray}
		\end{subequations}
		These are analogous to the ones obtained in Ref.~\cite{Barros}, with the variation that the scalar constraint \eqref{scal1} has been expressed in an alternative manner thanks to the definition
		\begin{equation}
		\label{J}
		J_{ai} := \left(P^{-1}\right)^{j}{}_{i} \left[\left(1-\frac{\sigma}{\gamma^{2}}\right) \Upsilon_{aj} - \frac{\sigma}{\gamma^{2}}A_{aj}\right],
		\end{equation}
		with $\Upsilon_{ai}$ being written in terms of the above canonical variables with the help of Eqs.~\eqref{solPsi} and \eqref{transfw}--\eqref{transfY}. For the sake of completeness, we write down below the expression for the diffeomorphism constraint~\eqref{DC} in terms of these variables,
		\begin{equation}
			\tilde{\mathcal{D}}_a=\dfrac{2}{\gamma}\tilde{E}^{bi}\partial_{[b}A_{a]i}+ \dfrac{1}{\gamma} A_{ai}\partial_b\tilde{E}^{bi}- \dfrac{1}{\gamma}\tilde{\zeta}_i\partial_a\chi^i,
		\end{equation}
		which establishes that under spatial diffeomorphisms $A_{ai}$ transforms as a 1-form, $\tilde{E}^{ai}$ as a vector density, $\tilde{\zeta}_i$ as a scalar density, and $\chi^i$ as a scalar function.

	\section{Time gauge}
		\label{TG}
		
		Up to now the Hamiltonian formulations contained in Secs.~\ref{Sol} and \ref{CV} are covariant under the full group $SO(\sigma)$; because of the splitting of the internal group into boosts and rotations, they might not display it manifestly, but they are indeed (see Ref.~\cite{MontRomCel,*MontRomEscCel} for a manifestly Lorentz-covariant approach). In order to make contact with the Ashtekar-Barbero formalism, we must impose the time gauge $\chi^{i}=0$ (assumed throughout this section), which removes the boost freedom of the theory. Regardless of which of the previous formulations we take as the starting point, this gauge condition together with the boost constraint $\tilde{\mathcal{G}}_{\mbox{{\tiny boost}}}^{i}$ form a second-class set, indicating that they both have to be solved jointly to successfully fix the boost freedom.
		
		From Eq.~\eqref{spamet} we observe that, in the time gauge, the variables $\tilde{E}^{ai}$ become the inverse of the densitized triad for the spatial submanifold $\Sigma$, whereas Eq.~\eqref{CD} implies that
		\begin{equation}
			\Gamma_{ai} := -\dfrac{1}{2} \epsilon_{ijk}\Gamma_{a}{}^{jk}\label{Gammaai}
		\end{equation}
		is the spin connection compatible with the densitized triad $\tilde{E}^{ai}$ [or the spatial Levi-Civita connection as seen from the $SO(3)$ frame], its expression being given by
		\begin{equation}
			\label{GS}
			\Gamma_{ai} = \epsilon_{ijk}\tilde{E}^{bj}\left( \partial_{[b}\underaccent{\tilde}{E}_{a]}{}^{k} + \underaccent{\tilde}{E}_{a}{}^{[l|}\tilde{E}^{c|k]}\partial_{b} \underaccent{\tilde}{E}_{cl} \right) .
		\end{equation}
		In the following paragraphs, we elaborate on the consequences of imposing the time gauge starting independently from the results of Secs. \ref{Sol} and \ref{CV}.

		\subsection{Time gauge in noncanonical coordinates}
		
			Looking first at the noncanonical approach of Sec.~\ref{Sol}, the solution of the boost constraint \eqref{Gboost} gives
			\begin{equation}
				\tilde{Y}_{i} = \dfrac{\sigma}{2\gamma}  \epsilon_{ijk} \omega_{a0}{}^{j} \tilde{E}^{ak},\label{Ytg}
			\end{equation}
			where we have made use of \eqref{CD} to cancel the terms involving the spin connection. Using the previous expression together with $\chi^{i}=0$ in the action~\eqref{S2} yields
			\begin{equation}
			S =  \int_\mathbb{R} dt \int_{\Sigma} d^{3}x \left( \mu_{ai} \dot{\tilde{E}}^{ai} + \tilde{\alpha}^{ai} \dot{\omega}_{a0i} - \tilde{H}' + \partial_{a}{\tilde{B}^{a}} \right), \label{actnon}
			\end{equation}
			in which, from Eqs.~\eqref{mult1} and~\eqref{mult3}, $\mu_{ai}$ and $\tilde{\alpha}^{ai}$ are given by 
			\begin{eqnarray}
			\mu_{ai} & := &  - \dfrac{2\sigma}{\gamma^{2}} \omega_{b0[i}\tilde{E}^{b}{}_{j]}\underaccent{\tilde}{E}_{a}{}^{j},\label{muai}  \\
			\tilde{\alpha}^{ai} & = & -2 \tilde{E}^{ai} \label{ai},
			\end{eqnarray}
			whereas the boundary term \eqref{bound} collapses  to
			\begin{equation}
				\tilde{B}^{a}=  \dfrac{1}{\gamma}\epsilon_{ijk} \underaccent{\tilde}{E}_{b}{}^{i}\tilde{E}^{aj}\dot{\tilde{E}}^{bk}.\label{Ba}
			\end{equation}
			In terms of the phase-space variables $(\omega_{a0i},\tilde{E}^{ai})$ the constraints \eqref{Grot}--\eqref{scal}, which make up $\tilde{H}'$, can be expressed as
			\begin{subequations}
				\begin{eqnarray}
					\tilde{\mathcal{G}}^{i} & = & \left(1 - \dfrac{\sigma}{\gamma^{2}}\right) \epsilon^{i}{}_{jk} \omega_{a0}{}^{j} \tilde{E}^{ak}, \label{gaussnon}\\
					\tilde{\mathcal{V}}_{a} & = & 2 \nabla_{[a}\left(\omega_{b]0i}\tilde{E}^{bi}\right)  + \frac{\sigma}{2(\gamma^2-\sigma)}\epsilon_{ijk}\tilde{E}^{bi}\underaccent{\tilde}{E}_{a}{}^{j}\nabla_{b}\tilde{\mathcal{G}}^{k},\notag \\ \label{vecnon}\\
					\tilde{\tilde{\mathcal{H}}} & = & \dfrac{\sigma}{2} \epsilon_{ijk}\tilde{E}^{ai}\tilde{E}^{bj} R_{ab}{}^{k} + \tilde{E}^{a[i|} \tilde{E}^{b|j]} \omega_{a0i}\omega_{b0j} \notag \\ 
					& & - \dfrac{\sigma \gamma^{2}}{4\left(\gamma^{2}-\sigma\right)^{2}}\tilde{\mathcal{G}}^{i}\tilde{\mathcal{G}}_{i} + \sigma \Lambda |\tilde{\tilde{E}}| ,\label{scalanon}
				\end{eqnarray}
			\end{subequations}
			where we have omitted the label ``rot'' in the rotational Gauss constraint, $R_{abi}:=2 \partial_{[a}\Gamma_{b]i}+\epsilon_{ijk}\Gamma_{a}{}^{j}\Gamma_{b}{}^{k}$ is the curvature of the spin connection $\Gamma_{ai}$ and $\nabla_a$ is the full covariant derivative associated to it, that is, $\nabla_a$ annihilates $\tilde{E}^{ai}$ [see Eq.~\eqref{CD}],
			\begin{equation}
				\nabla_a\tilde{E}^{bi}:=\partial_{a} \tilde{E}^{bi} + \Gamma^{b}{}_{ca}\tilde{E}^{ci} - \Gamma^{c}{}_{ca} \tilde{E}^{bi} + \epsilon^{i}{}_{jk}\Gamma_{a}{}^{j}\tilde{E}^{bk}  = 0.
			\end{equation}
			Notice that in Eq. \eqref{vecnon} the variables $\omega_{a0i}$ are treated as spatial 1-forms by $\nabla_a$, whereas $\tilde{\mathcal{G}}^{i}$ is considered as a densitizated internal vector (for instance, $\nabla_{a}\tilde{\mathcal{G}}^{i} = \partial_{a}\tilde{\mathcal{G}}^{i} - \Gamma^{b}{}_{b a} \tilde{\mathcal{G}}^{i} + \epsilon^{i}{}_{jk} \Gamma_{a}{}^{j}\tilde{\mathcal{G}}^{k}$). By comparing the Gauss constraint \eqref{gaussnon} with Eq.~\eqref{muai} we immediately conclude that
			\begin{equation}
				\mu_{ai}=-\frac{\sigma}{\gamma^2-\sigma}\epsilon_{i j k}\underaccent{\tilde}{E}_{a}{}^{j}\tilde{\mathcal{G}}^{k}\approx 0,
			\end{equation}
			which means that, in Eq.~\eqref{actnon}, the variables $(\omega_{a0i},\tilde{E}^{ai})$ are actually canonical, since we can eliminate the first term of the integrand of the action~\eqref{actnon} by redefining the Lagrange multiplier that accompanies the Gauss constraint inside $\tilde{H}'$. Notice that the set of constraints~\eqref{gaussnon}--\eqref{scalanon} resembles that of the $SO(3)$ ADM formalism, which arises from the canonical analysis of the Palatini action in the time gauge~\cite{ashtekar1991lectures}. Since this action corresponds to the limit $\gamma^{-1}\rightarrow 0$ of the Holst action, the constraints~\eqref{gaussnon}--\eqref{scalanon} in that limit indeed reproduce those of the $SO(3)$ ADM formalism ($-\omega_{a0i}$ gets identified with the extrinsic curvature). Even more remarkably, although the above constraints explicitly depend on the Immirzi parameter, it appears as a global factor in Eq.~\eqref{gaussnon} and as a multiplicative factor of the terms proportional to $\tilde{\mathcal{G}}^{i}$ in the other two constraints. We can then rescale the Gauss constraint and appeal again to the redefinition of the Lagrange multiplier appearing in front of the Gauss constraint in  $\tilde{H}'$ to cast the action in such a way that the Immirzi parameter does not explicitly show up in it (because of the term with spatial derivatives of the Gauss constraint in the vector constraint, there are boundary terms involved in the process, but they can be dropped if the spatial 3-manifold has no boundary). Thus, the Immirzi parameter still remains classically undetectable in the action~\eqref{actnon}, and the constraints~\eqref{gaussnon}--\eqref{scalanon} actually correspond to the $SO(3)$ ADM formalism~\cite{ashtekar1991lectures} after making the redefinitions already explained.
			

			We now make contact with the Ashtekar-Barbero formulation. We point out that, in order to arrive exactly at the same results as in the following subsection, we will not neglect the first term of the integrand of Eq.~\eqref{actnon} but rather it will be included in the definition of the new phase-space variables; if we decide to disregard it, the resulting sets of constraints then differ from each other by terms proportional to the Gauss constraint (see below). First, from Eqs. \eqref{muai} and \eqref{Ba}, we obtain the following identity:
			\begin{equation}
				\mu_{ai} \dot{\tilde{E}}^{ai} + \partial_{a}{\tilde{B}^{a}}=\dfrac{2}{\gamma}\tilde{E}^{ai}\partial_t\left(\Gamma_{ai} - \frac{\sigma}{\gamma} \underaccent{\tilde}{E}_{a}{}^{j}\tilde{E}^b{}_{[i|}\omega_{b0|j]}\right).
			\end{equation}
			Using this, the action \eqref{actnon} takes the form
			 \begin{eqnarray}
				\label{STG}
				S & = & \int_\mathbb{R} dt \int_{\Sigma} d^{3}x \left[ \frac{2}{\gamma} \tilde{E}^{ai}\dot{A}_{ai} - \tilde{H}' \right], 
			\end{eqnarray}
			whereof it is clear that the variables $\tilde{E}^{ai}$ and
			\begin{eqnarray}
				\label{A}
				A_{ai}:= -\gamma \omega_{a0i} + \Gamma_{ai} - \frac{\sigma}{\gamma} \underaccent{\tilde}{E}_{a}{}^{j}\tilde{E}^b{}_{[i|}\omega_{b0|j]},\label{Abarb}
			\end{eqnarray}
			are canonically conjugate to each other, then satisfying $\{ A_{ai}(x),\, \tilde{E}^{bj}(y)\}=(\gamma/2)\delta^{b}_{a}\delta^{j}_{i}\delta^{3}(x-y)$. The change of variables~\eqref{Abarb} resembles Barbero's canonical transformation except for the last term, which is the contribution coming from $\mu_{ai}$ (which---we remind the reader---is proportional to the Gauss constraint). This is the same expression one finds after imposing the time gauge in Eq.~\eqref{transfw}, and that is why we decided to preserve the term proportional to the Gauss constraint and denoted by the same symbol $A_{ai}$ to the new configuration variables. It is worth emphasizing the usefulness of the time gauge not only for rendering the complicated symplectic structure of Sec.~\ref{Sol} canonical in the initial variables $(\omega_{a0i},\tilde{E}^{ai})$, but also for helping to uncover a canonical transformation linking these variables with $(A_{ai},\tilde{E}^{ai})$. 
			
			To express the constraints \eqref{gaussnon}--\eqref{scalanon} in terms of the canonical pair $(A_{ai},\, \tilde{E}^{ai})$ we first have to invert Eq.~\eqref{Abarb} for $\omega_{a0i}$, obtaining
			\begin{equation}
				\omega_{a0i} =  \dfrac{1}{2\gamma} \Bigg[ \dfrac{2\gamma^{2}-\sigma}{\gamma^{2}-\sigma} \delta^{j}_{i} \delta^{b}_{a} - \dfrac{\sigma}{\gamma^{2}-\sigma}\tilde{E}^{b}{}_{i} \underaccent{\tilde}{E}_{a}{}^{j} \Bigg] \left( \Gamma_{b j} - A_{b j} \right).
			\end{equation}
			Plugging this back into Eqs.~\eqref{gaussnon}--\eqref{scalanon} leads to
			\begin{subequations}
				\label{constTG}
				\begin{eqnarray}
					\tilde{\mathcal{G}}^{i} &=& - \dfrac{1}{\gamma}\left(\partial_{a} \tilde{E}^{ai} + \epsilon^{ijk} A_{aj} \tilde{E}^{a}{}_{k}\right),\label{constTG1}\\
					\tilde{\mathcal{V}}_{a} & = & \dfrac{1}{\gamma}\tilde{E}^{bi}F_{bai} +  \left(\Gamma_{ai} -  A_{ai}\right)\tilde{\mathcal{G}}^{i}, \\
					\tilde{\tilde{\mathcal{H}}} & = & \dfrac{1}{2 \gamma^{2}}\epsilon_{ijk}\tilde{E}^{ai}\tilde{E}^{bj} \left[ F_{ab}{}^{k} + \left( \sigma \gamma^{2}-1 \right) R_{ab}{}^{k} \right] \notag \\
						& & -\frac{1}{\gamma} \tilde{E}^{a}{}_i\nabla_a \tilde{\mathcal{G}}^i +\dfrac{\sigma}{4\left(\gamma^{2}-\sigma\right)} \tilde{\mathcal{G}}^{i}\tilde{\mathcal{G}}_{i} + \sigma \Lambda |\tilde{\tilde{E}}|,\label{constTG2}
				\end{eqnarray}
			\end{subequations}
			with $F_{abi}:=2 \partial_{[a} A_{b]i}+\epsilon_{ijk}A_{a}{}^{j} A_{b}{}^k$ being the strength of the connection $A_{ai}$. 
			
			If, on the other side, we decide to neglect the third term on the right-hand side of Eq.~\eqref{Abarb} and work with the Barbero's original connection ${}_{B}A_{ai}:= -\gamma \omega_{a0i} + \Gamma_{ai}$ (${}_{B}F_{abi}:=2 \partial_{[a|} {}_{B}A_{|b]i}+\epsilon_{ijk}{}_{B}A_{a}{}^{j} {}_{B}A_{b}{}^k$), the constraints become
			\begin{subequations}
				\label{constTGAB}
				\begin{eqnarray}
				\vspace{-5mm}\tilde{\mathcal{G}}^{i} &=& - \dfrac{(\gamma^2-\sigma)}{\gamma^3}\left(\partial_{a} \tilde{E}^{ai} + \epsilon^{ijk} {}_{B}A_{aj} \tilde{E}^{a}{}_{k}\right),\label{constTG1AB}\\
				\tilde{\mathcal{V}}_{a} & = & \dfrac{1}{\gamma}\tilde{E}^{bi}{}_{B}F_{bai}+ \frac{\sigma}{2(\gamma^2-\sigma)}\epsilon_{ijk}\tilde{E}^{bi}\underaccent{\tilde}{E}_{a}{}^{j}\nabla_{b}\tilde{\mathcal{G}}^{k}\notag\\
				&& +  \frac{\gamma^ 2}{\gamma^2-\sigma}\left(\Gamma_{ai} -  {}_{B}A_{ai}\right)\tilde{\mathcal{G}}^{i}, \\
				\tilde{\tilde{\mathcal{H}}} & = & \dfrac{1}{2 \gamma^{2}}\epsilon_{ijk}\tilde{E}^{ai}\tilde{E}^{bj} \left[ {}_{B}F_{ab}{}^{k} + \left( \sigma \gamma^{2}-1 \right) R_{ab}{}^{k} \right] \notag \\
				& & -\frac{\gamma}{\gamma^2-\sigma} \tilde{E}^{a}{}_i\nabla_a \tilde{\mathcal{G}}^i -\dfrac{\sigma\gamma^2}{4\left(\gamma^{2}-\sigma\right)} \tilde{\mathcal{G}}^{i}\tilde{\mathcal{G}}_{i} + \sigma \Lambda |\tilde{\tilde{E}}|,\notag\\
				\label{constTG2AB}
				\end{eqnarray}
				\end{subequations}			
			which take exactly the same form as Eqs.~\eqref{constTG1}--\eqref{constTG2} except for the global factor in the expression of the Gauss constraint and the constant factors in front of the terms proportional to $\tilde{\mathcal{G}}^{i}$ in the remaining constraints. Both sets of constraints, Eqs.~\eqref{constTG1}--\eqref{constTG2} or Eqs.~\eqref{constTG1AB}--\eqref{constTG2AB}, embody the Ashtekar-Barbero formulation.

		\subsection{Time gauge in Darboux coordinates}

			If, on the other side, we start from the canonical formulation of Sec.~\ref{CV}, the implementation of the time gauge in the action~\eqref{S3} immediately leads to the Ashtekar-Barbero formulation. To get this, we just have to note two things; first, that in the time gauge the variable $J_{ai}$ becomes
			\begin{eqnarray}
				J_{ai} &=& -\dfrac{1}{2} \left( \delta^{b}_{a} \delta^{j}_{i} + \tilde{E}^{b}{}_{i} \underaccent{\tilde}{E}_{a}{}^{j}\right)\left[\dfrac{\sigma}{\gamma^{2}}A_{bj} + \left(1- \dfrac{\sigma}{\gamma^{2}}\right)\Gamma_{bj}\right] \notag \\
				\label{JTG}
				& & + \dfrac{1}{2\gamma}\epsilon_{ijk} \underaccent{\tilde}{E}_{a}{}^{j} \tilde{\zeta}^{k},
			\end{eqnarray}
			and second, that we need the solution of the boost constraint~\eqref{boostCV}, which in these variables reads
			\begin{equation}
				\tilde{\zeta}_{i} =  \gamma \partial_{a}\tilde{E}^{a}{}_{i} = -\gamma \epsilon_{ijk}\Gamma_{a}{}^{j}\tilde{E}^{ak},
			\end{equation}
			where, to obtain the last equality, we have used Eqs.~\eqref{CD} and \eqref{Gammaai}. Substituting both expressions in Eqs.~\eqref{gaussCV}--\eqref{scal1} (together with $\chi^{i}=0$) readily collapses the latter into Eqs.~\eqref{constTG1}--\eqref{constTG2} without any further consideration.

	\section{Discussion}
		\label{Concl}
	
		In this paper we have solved, without resorting to the time gauge, the second-class constraints arising in the Lorentz-covariant canonical analysis of the Holst action with a cosmological constant. We have followed a path closely related to that of Cianfrani and Montani~\cite{Cianfraniprl}, finding a complete parametrization of the phase space in terms of the 24 noncanonical coordinates ($\tilde{E}^{ai}, \, \chi_{i}, \, \omega_{a0i}, \,\tilde{Y}^{i}$) subject to the ten first-class constraints \eqref{Gboost}--\eqref{scal}, which means that the reduced phase space of the ensuing theory has dimension four at each spatial point, as expected for general relativity. We then performed a Darboux transformation that allowed us to establish the right link with the canonical formulation of Barros~\cite{Barros}. At the end we showed, in both the noncanonical and the canonical parametrizations of the phase space, how the Ashtekar-Barbero variables are obtained once the time gauge is implemented.

		Our main motivation to do this work was to find the missing link between the canonical formulation presented by Cianfrani and Montani, and the one reported by Barros. It all stems from the fact that the solution of the second-class constraint~\eqref{psi} given by the former authors corresponds to a particular solution of the associated system of equations and not to its general solution. Indeed, the solution given in Ref.~\cite{Cianfraniprl} neglects part of the solution of the homogeneous system, thus not only providing an incomplete solution, but also an insufficient number of independent variables to properly label the points of the kinematic phase space since the variables $\tilde{Y}^{i}$ were completely ignored in their approach. In consequence, Cianfrani and Montani's solution is not enough to establish an appropriate link with Barros's formulation, which actually gives a correct description of the phase space of general relativity in terms of canonical variables. Hence, in this paper we have healed the mismatch between both approaches, first by solving correctly the aforementioned second-class constraints, and later by providing the invertible Darboux transformation connecting our formulation with Barros's one. In addition, we have provided explicit expressions of the constraints in terms of the noncanonical variables ($\tilde{E}^{ai}, \, \chi_{i}, \, \omega_{a0i}, \,\tilde{Y}^{i}$) [see Eqs.~\eqref{Gboost}--\eqref{scal}], which now become the starting point before any attempt to implementing a determinate gauge fixing on this formulation.
		
		On the other side, since Cianfrani and Montani omitted the parameters $\tilde{Y}^{i}$, their approach to exhibit the $SU(2)$ invariance of the theory with a nondynamical $\chi^{i}$ (and even with a dynamical one) must be reconsidered. Assuming---as they do in the first part of their analysis---that $\chi^{i}$ is time independent ($\dot{\chi}^{i}=0$) does not mean that $\tilde{Y}^{i}$ has to vanish, but rather that it must be fixed by the simultaneous solution of an appropriate combination of boosts and rotations compatible with the assumed condition on $\chi^{i}$ (they have to form a second-class set). In this regard, the work of Ref.~\cite{Liu_Noui}, where the authors perform a gauge fixing along a nondynamical $\chi^{i}$ using Barros' parametrization, may help to work out the same kind of gauge fixing in the variables found at the end of Sec.~\ref{Sol} and subsequently allow us to write the resulting theory in an explicitly $SU(2)$ or $SU(1,1)$ invariant fashion with respect to the fixed $\chi^{i}$.

		Although it is true that the phase-space variables involved in the constraints \eqref{Gboost}--\eqref{scal} have associated a complicated and noncanonical symplectic structure, it is remarkable how neatly the Ashtekar-Barbero formulation is obtained from them in the time gauge. The link between the formulation of this paper and the one due to Barros---something that is made tangible thanks to the Darboux transformation~\eqref{transf1}--\eqref{transf2}---actually allows us to express the analogues of the constraints of Barros (in particular the form of the scalar constraint), namely Eqs.~\eqref{boostCV}--\eqref{scal1}, in such a way that by enforcing the time gauge there we are immediately led to the Ashtekar-Barbero variables too. In constrast, the derivation of the latter in Barros's work follows a different and longer path where the expressions of the resulting constraints are not used directly, but rather he goes back to the solution of the second-class constraints and solves them jointly with the restraints entailed by the time gauge. We stress that in our approach this is not necessary: we can impose the time gauge directly on the constraints~\eqref{boostCV}--\eqref{scal1} and arrive at the Ashtekar-Barbero formulation without any effort.
		
		Finally, it would also be interesting to establish a link between the formulations contained in this paper and the manifestly Lorentz-covariant ones of Ref.~\cite{MontRomCel,*MontRomEscCel}. After all, the canonical theories with nonmanifest Lorentz symmetry have enriched the discussion around the significance of the time gauge in quantum gravity (this was also the idea of Ref.~\cite{Cianfraniprl}, even if their approach is incomplete), allowing the identification of Lorentz-covariant connection variables and fluxes with interesting results as to the relevance of the Immirzi parameter at the quantum level~\cite{NouiSIGMA72011,Nouiprd84044002}. We can then ask whether the same kind of information can be extracted from the Lorentz-covariant variables of Ref.~\cite{MontRomCel,*MontRomEscCel} while preserving the manifestly Lorentz invariance of the theory. Perhaps, this could enlighten the path towards solving the ambiguities posed by the Immirzi parameter once for all.

 	\section*{Acknowledgments}
 	
 	This work was partially supported by Fondo SEP-Cinvestav and by Consejo Nacional de Ciencia y Tecnolog\'{i}a (CONACyT), M\'{e}xico, Grants No.~237004-F and No.~A1-S-7701.

	\bibliography{references}

\end{document}